\documentclass[twocolumn,showpacs,unsortedaddress,prl]{revtex4}

\usepackage{amsfonts,amssymb,amsmath}
\usepackage{bm}

\usepackage[OT1]{fontenc}

\usepackage{graphicx}

\newcommand{\Na}[1]{Na$_{#1}$CoO$_2$}

\newcommand{\vect}[1]{\bm{#1}}

\newcommand{\moy}[1]{\langle #1 \rangle}

\newcommand{\bra}[1]{\langle #1 \rvert}
\newcommand{\ket}[1]{\lvert #1 \rangle}

\begin{document}

\title{Electronic structure and Fermi surface tolopogy of \Na{x}}

\author{A.~Bourgeois}
\affiliation{Laboratoire de Physique des Solides, CNRS UMR-8502, Universit\'e Paris XI, 91405 Orsay cedex, France}
\author{A.A.~Aligia}
\affiliation{Comisi\'on Nacional de Energ\'ia At\'omica, Centro At\'omico Bariloche and Instituto Balseiro, 8400 S.C. de Bariloche, Argentina}
\author{T.~Kroll}
\affiliation{IFW Dresden, P.O. Box 270016, D-01171 Dresden, Germany}
\author{M.D.~N\'u\~nez-Regueiro}
\affiliation{Laboratoire de Physique des Solides, CNRS UMR-8502, Universit\'e Paris XI, 91405 Orsay cedex, France}


\date{\today}

\begin{abstract}

We construct an effective Hamiltonian for the motion of $T_{2g}$ highly correlated 
states in \Na{x}. We solve exactly a multiband model in a CoO$_{6}$ cluster
with electronic occupation corresponding to a nominal Co valence of either $+3$ or $+4$. 
Using the ensuing ground states, we calculate the effective O mediated hopping $t = 0.10$~eV 
between many-body $T_{2g}$ states, and estimate the direct hopping $t' \sim 0.04$~eV. 
The trigonal splitting $3D = 0.315$~eV is taken from recent quantum chemistry calculations. 
The resulting effective Hamiltonian is solved using a generalized slave-boson mean-field 
approximation. The results show a significant band renormalization and a Fermi surface 
topology that agrees with experiment, in contrast to predictions using the local-density 
approximation.

\end{abstract}


\maketitle

The doped layered hexagonal cobaltates \Na{x} have attracted considerable interest in the 
last few years. The exceptionally high thermopower and at the same time low thermal 
conductivity and resistivity for $0.5 < x < 0.9$~\cite{PRB65195106,JJAP427383}, makes the 
system attractive because of its potential technological applications. 
Wider attention has been triggered by the discovery of superconductivity in hydrated 
\Na{0.35}--1.3~H$_{2}$O~\cite{NAT42253}. Because of the hexagonal structure of the 
system which frustrates competing magnetic interactions, a possible natural explanation of the 
superconductivity seemed to be a resonance-valence-bond state in an effective one-band 
model~\cite{PRL91097003}. However, based on geometrical and physical arguments, Koshibae 
and Maekawa pointed out that it is more realistic to start from a three-band model of 
degenerate Co $3d$ $t_{2g}$ orbitals with an indirect hopping over intermediate O 
orbitals~\cite{PRL91257003,PRB71214414}. This leads to four independent interpenetrating 
Kagom\'e sublattices. On the other hand, local-density approximation (LDA) calculations 
predict a Fermi surface (FS) with a large portion around the $\varGamma$ 
point of the Brillouin zone, and six pockets near the $K$ points~\cite{PRB6113397,PRB70045104}. 
However, angle-resolved photoemission (ARPES) revealed only a single hole like 
FS around $\varGamma$ and no hole pockets~\cite{PRL92246402,PRL92246403,PRL95146401}. 
Recent experiments with X-ray absorption spectroscopy (XAS) ~\cite{PRL94146402,PRB74115123}
and their interpretation~\cite{PRL94146402,PRB74115123,PRB74115124} show a strong Co--O covalency. 
In particular, the amount of the $3d^{n}$ configuration in a CoO$_{6}$ cluster is smaller than the total 
amount of $3d^{n+1} L$ states ($L$ denotes a ligand O $2p$ hole) for both, nominal 
Co$^{3+}$ ($n = 6$) and Co$^{4+}$ ($n = 5$). Ab initio quantum chemical calculations also obtain a large
Co--O hybridization~\cite{CM0605454}.

It has been suggested that the discrepancy between the LDA calculations and the Fermi
surface measured by ARPES can be explained as an effect of the correlations ~\cite{PRL94206401},
which are absent in LDA and LDA+U approaches, even in the most sophisticated treatments~\cite{PRL93236402}. 
However, the effective Hamiltonian used is unjustified and differs from that we have derived (see below). 
Moreover, the dispersion relation along $\Gamma$-K differs from the experimentally observed one, and the 
results contradict those obtained using dynamical mean field theory,
which actually obtain an enhancement of the hole pockets~\cite{PRL94196401}. Another recently
proposed explanation is a localization of the hole pockets by disorder~\cite{CM0604002}. 
However this localization is not expected to alter the small volume enclosed by the main 
portion of the LDA Fermi surface around $\varGamma$, while according to the more recent
ARPES experiments, this portion contains all holes in a way consistent with Luttinger
theorem~\cite{PRL95146401}.

This discussion shows the need of further studies of the electronic structure of \Na{x}, 
and the starting point should take into account the strong Co--O covalency. 
This is the purpose of this Letter. We start from the exact solution of the appropriate 
multiband Hamiltonian for Co $3d$ and O $2p$ holes in the basic CoO$_{6}$ octahedron, 
with the hole content corresponding to $x = 1$ (nominal Co$^{3+}$) and with one additional 
hole (nominal Co$^{4+}$ as for $x = 0$). The parameters of the multiband model are 
fixed by a previous fitting of the polarization dependent 
XAS spectra~\cite{PRB74115124}. The resulting ground states are mapped onto the 
corresponding states of an effective three-band model for (fictitious) $3d$ $t_{2g}$ 
orbitals on Co sites. The eigenstates of the cluster are used to calculate the hopping in 
the effective model. The procedure has some similarities to the derivation of an effective 
$t-J$ model in the cuprates using non-orthogonal 
Zhang-Rice singlets~\cite{PRB397375,PRB4913061}. The trigonal splitting is taken 
from recent difference-dedicated configuration interaction (DDCI) calculations~\cite{CM0605454}
in a distorted CoO$_{6}$ cluster. The DDCI are quantum chemistry calculations,
particularly suited to calculate excitation energies in strongly 
correlated systems~\cite{CM0605454,CPL238222}.   
The effective model is solved in a slave-boson mean-field
approximation~\cite{PRL571362,PRB5612909JPSJ661391PRB475095}.
The results reproduce the essential results of ARPES, in 
particular the shape of the FS and the quasiparticle dispersion
near the Fermi energy $\epsilon_F$.

It might seem that the procedure mentioned above, 
which eliminates the O $2p$ states from the low-energy physics, 
is not valid when the Co--O hopping is larger or of the 
order of the Co and O on-site energy difference $\Delta$. 
However, calculations in the cuprates~\cite{PRB538751,PRB565637} 
and other systems~\cite{PRB66104418} using the cell perturbation method 
showed that this is not the case. In fact as the hopping 
increases, the $d$ states are replaced by hybrid states of the same 
symmetry which remain well separated from the other states. 
Using the mapping procedure it is even possible to calculate the O spectral density at low 
energies with the effective model which does not contain explicitly O states~\cite{PRB565637}.

The Hamiltonian for the CoO$_{6}$ cluster contains Co $3d$ and O 
$2p$ holes and includes the splitting between $3d$ $e_{g}$ and $t_{2g}$ 
orbitals, Co--O and O--O hoppings and all 
interactions between $3d$ holes. Spin-orbit interaction is neglected. 
Details are given in Ref.~\cite{PRB74115124}. We assume octahedral $O_{h}$ 
symmetry. This assumption greatly reduces the size of the matrices to be 
diagonalized and has a very small effect on the resulting many-body 
eigenstates~\cite{PRB74115124}. 
The ground state of the cluster with four holes (nominal Co$^{3+}$) 
$\ket{A_{1g}}$ is a singlet with $A_{1g}$ symmetry. We use capital 
letters to denote the irreducible representations of many-body states. 
For the parameters that fit the polarization dependent XAS spectra,
it has been found that $\ket{A_{1g}}$ consists mainly of 30\% of the state with 
four $3d$ $e_{g}$ holes and 47\% of states with three $3d$ $e_{g}$ holes 
and one $2p$ hole, in a linear combination of $e_{g}$ symmetry ~\cite{PRB74115124}. 
For one additional hole, the ground state of the cluster is a spin doublet with 
$T_{2g}$ symmetry (six-fold degenerate). These states will be denoted as 
$\ket{\gamma,\sigma}$, where $\gamma = xy, yz \text{ or } zx$ and 
$\sigma = \uparrow \text{or} \downarrow$. As a first approximation, 
they can be constructed adding a $3d$ $t_{2g}$ hole to $\ket{A_{1g}}$. 

\begin{figure}[htbp]
	\includegraphics[width=.7\linewidth]{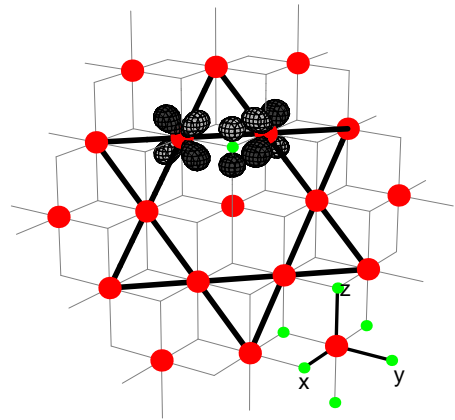}
	\caption{(Color online) Orbitals involved in the effective hopping of a hole with symmetry $yz$ to a neighboring state with symmetry $zx$, through a $p_{z}$ orbital of an intermediate O site.}
	\label{Fig:Hopping}
\end{figure}

In this work, to represent the movement of holes in the system between different 
CoO$_{6}$ clusters, we map the state $\ket{i,A_{1g}}$ where $i$ is a 
site index, onto the vacuum (absence of holes) at site $i$, and the 
states $\ket{i,\gamma,\sigma}$ onto the corresponding states 
$h^{\dagger}_{i\gamma\sigma} \ket{0}$ of the effective three-band model, 
where $h^{\dagger}_{i \gamma \sigma}$ creates a hole at site $i$, 
orbital $\gamma$ and spin $\sigma$. The most important effective hopping 
is the one mediated by O orbitals. Geometrical considerations show that 
a hole in the state $\ket{i,\gamma,\sigma}$ can hop only to some nearest 
Co atoms (building the state $\ket{j,\gamma',\sigma}$) through an 
intermediate O $2p$ orbital. Fig.~\ref{Fig:Hopping} illustrates this 
indirect hopping, which can be well approximated by~\cite{PRB4913061}
\begin{equation}
	-t = \bra{j,zx,\uparrow} \bra{i,A_{1g}} H_{hop} \ket{i,yz,\uparrow} \ket{j,A_{1g}}
	\label{Eq:t}
\end{equation}
$H_{hop}$ is the sum of Co--O and O--O hopping terms 
of both clusters. For the parameters that fit the XAS spectra we obtain 
$t = 0.10$~eV. This agrees with previous estimations~\cite{PRL91257003}.
Considering all possible indirect hoppings leads to the 
picture of four interpenetrating Kagom\'e sublattices for the degenerate 
$\ket{\gamma,\sigma}$ orbitals (see Refs.~\cite{PRL91257003,PRB71214414} and Fig.~\ref{Fig:Hopping}).

While $t$ is the more important hopping process~\cite{PRB71214414}, 
the direct hopping $t_{1}$ between $d$ electrons might be important for 
a realistic description of the band structure~\cite{PRL91257003,PRL94206401}. 
Unfortunately, since this hopping is not included in the calculation of 
the CoO$_{6}$ cluster used to fit the XAS spectra, it cannot be extracted 
using this experimental information. Taking the value $t_{1} = -44.6$~meV 
obtained from a fit of the band structure in Ref.~\cite{PRL94206401}, we 
obtain $t' = 36 \text{ meV} > 0$ for the effective hopping between highly 
correlated hole states
\begin{equation}
	t' = \bra{j,xy,\uparrow} \bra{i,A_{1g}} -t_{1} d^{\dagger}_{j,xy,\uparrow} d^{}_{i,xy,\uparrow} \ket{i,xy,\uparrow} \ket{j,A_{1g}}
	\label{Eq:t'}
\end{equation}
Here $i$ and $j$ refer to two nearest-neighbor clusters in the $(1,1,0)$ 
direction.

In presence of a trigonal distortion, the point symmetry is reduced to 
$D_{3d}$ and the $T_{2g}$ states are split into $A'_{1g}$ and $E'_{g}$ 
states. The prime denotes irreducible representations of $D_{3d}$. 
This splitting, which is an essential parameter of the effective model, 
is beyond the reach of our cluster calculation in $O_{h}$ symmetry. 
Fortunately, recent accurate DDCI calculations give a splitting 
$3D = 0.315$~eV between the many-body ground state 
$(\ket{i,xy,\sigma} + \ket{i,yz,\sigma} + \ket{i,zx,\sigma}) / \sqrt{3}$
of $A'_{1g}$ symmetry and the excited $E'_{g}$ states.
The large value of $D$ compared to estimations using
a point charge model~\cite{PRL91257003} is due to the effects
of the Co--O hybridization.

For $D \gg t$, the one-band model is justified, while for $D \ll t$, 
the picture based on four Kagom\'e sublattices is a better starting point
(Refs.~\cite{PRL91257003,PRB71214414} take $D = 0$). 
As explained above, the cluster calculations give $D \sim t$, which 
corresponds to an intermediate situation. This is also supported by an 
analysis of the XAS intensities under different polarizations of the 
incident light~\cite{PRB74115124}. In what follows we take 
$D = t$, $t' = 0.4\ t$ with $t \sim 0.1$~eV. Note that the indirect 
hopping of holes through O orbitals $-t$ and the direct hopping $t'$ have different 
signs, but we define them in such a way that $t > 0$, $t' > 0$.

The energies $t$, $D$ and $t'$ are smaller by at least one order of 
magnitude in comparison with all other energy scales in the problem, like 
Co--O charge-transfer energy $\Delta \sim 4$~eV and Coulomb 
repulsion at the same $d$ orbital 
$U \sim 4.5$~eV~\cite{PRL94146402,PRB74115124,CM0605454}. Therefore, we assume that 
the energy cost to bring two additional holes in the same cluster is much 
larger than the effective hopping. This implies that double occupancy is 
forbidden in the effective low-energy model.

This model can thus be written in the form
\begin{equation}
	\begin{split}
		H_{eff} &= H_{1} + H_{int} \\
			&= \sum_{i,j}\sum_{\gamma,\gamma',\sigma} \bigl( t_{ij\gamma\gamma'} + D_{\gamma\gamma'} \delta_{ij} \bigr) h^{\dagger}_{i\gamma\sigma} h^{}_{j\gamma'\sigma} \\
			&\quad + U \sum_{i} \sum_{\gamma,\sigma \neq \gamma',\sigma'} h^{\dagger}_{i\gamma\sigma} h^{}_{i\gamma\sigma} h^{\dagger}_{i\gamma'\sigma'} h^{}_{i\gamma'\sigma'}
	\end{split}
	\label{Eq:Heff}
\end{equation}
with $U \rightarrow \infty$ to avoid double occupancy. 
Fourier transforming the hole operators, $H_{1}$ takes the explicit form
\begin{gather}
	H_{1} = \sum_{\vect{k}} \sum_{\gamma,\gamma',\sigma} \big( \epsilon_{\gamma\gamma'}(\vect{k}) + \epsilon'_{\gamma\gamma'}(\vect{k})
		+ D_{\gamma\gamma'} \big) h^{\dag}_{\vect{k}\gamma\sigma} h^{}_{\vect{k}\gamma'\sigma} \notag \\
	\epsilon_{\gamma\gamma'}(\vect{k}) = -2t
		\begin{pmatrix}
		0 & \cos \theta_{3} & \cos \theta_{2} \\
		\cos \theta_{3} & 0 & \cos \theta_{1} \\
		\cos \theta_{2} & \cos \theta_{1} & 0
		\end{pmatrix} \notag \\
	\epsilon'_{\gamma\gamma'}(\vect{k}) = 2t'
		\begin{pmatrix}
		\cos \theta_{1} & 0 & 0 \\
		0 & \cos \theta_{2} & 0 \\
		0 & 0 & \cos \theta_{3}
		\end{pmatrix} \notag \\
	D_{\gamma\gamma'} = -D
		\begin{pmatrix}
		0 & 1 & 1 \\
		1 & 0 & 1 \\
		1 & 1 & 0
		\end{pmatrix} \label{Eq:Hhop}
\end{gather}
with $\theta_{1} = k_{x} a$, 
$\theta_{2} = \frac{k_{x} a}{2} - \frac{\sqrt{3} k_{y} a}{2}$ 
and $\theta_{3} = \frac{k_{x} a}{2} + \frac{\sqrt{3} k_{y} a}{2}$, 
where here $k_{x}$ refers to the direction of two nearest-neighbor 
Co atoms ($(-1,1,0)$ in Fig.~\ref{Fig:Hopping}). 
We believe that $H_{eff}$ is the starting point for the description of 
the low-energy physics of the cobaltates. While it is similar to previous 
proposals~\cite{PRL91257003,PRL94206401}, the most important parameters 
$D$ and $t$ are derived from cluster calculations that properly
take into account the important effects of Co--O covalency
and all correlations inside the Co 3d shell.
Due to the large Co--O hybridization, 
the effect of the distortion of the 
CoO$_{6}$ octahedra is stronger than in previous multiband approaches, 
but not so strong to justify a one-band model based on localized 
$a'_{1g}$ orbitals.

In the following, we test the ability of $H_{eff}$ with parameters fixed 
as explained above, to explain the ARPES experiments. We solve it using a 
generalization to the multiband case of the slave boson 
treatment of Kotliar and Ruckenstein in mean field, which is equivalent to 
the Gutzwiller approximation~\cite{PRL571362}. The three-band case of 
$t_{2g}$ orbitals with only one relevant spin~\cite{PRB66104418} and 
the general multiband case~\cite{PRB5612909JPSJ661391PRB475095} 
were considered before. The basic idea is to enlarge the Fock space to 
include bosonic states which correspond to each state in the fermionic 
description. The vacuum state at site $i$ is now represented as 
$e^{\dagger}_{i} \ket{0}$, a state with one hole as 
$s^{\dagger}_{i\gamma\sigma} h^{\dagger}_{i\gamma\sigma} \ket{0}$, 
a state with two holes as 
$d^{\dagger}_{i\gamma\sigma\gamma'\sigma'} h^{\dagger}_{i\gamma\sigma} h^{\dagger}_{i\gamma'\sigma'} \ket{0}$, 
etc., where $e^{\dagger}_{i}$, $s^{\dagger}_{i\gamma\sigma}$, 
$d^{\dagger}_{i\gamma\sigma\gamma'\sigma'}$\ldots are bosonic 
operators corresponding to empty, singly occupied, doubly occupied sites 
and so on. In our case, with $U \rightarrow \infty$, only the first two 
bosons are relevant and from now on we neglect the others in the description that follows to simplify it.
The boson operators should satisfy the constraints
\begin{equation}
	e^{\dagger}_{i} e_{i} + \sum_{i,\gamma,\sigma} s^{\dagger}_{i\gamma\sigma} s_{i\gamma\sigma} = 1; \quad
	h^{\dagger}_{i\gamma \sigma} h_{i\gamma\sigma} = s^{\dagger}_{i\gamma\sigma} s_{i\gamma\sigma}
	\label{Eq:Cons1}
\end{equation}
The first one states that there is only one boson at each site and
the second that counting states with one hole in the fermionic or 
bosonic representations should lead to the same result.
In the hopping terms involving different sites 
($i \neq j$ in Eq.\eqref{Eq:Heff}), 
the fermion creation operators should be accompanied by their respective 
bosonic part
\begin{equation}
	h^{\dagger}_{i \gamma\sigma} \rightarrow h^{\dagger}_{i\gamma\sigma}
		\bigl( 1 - e^{\dagger}_{i} e^{}_{i} \bigr)^{-\frac{1}{2}} s^{\dagger}_{i\gamma\sigma} e^{}_{i}
		\bigl( 1 - s^{\dagger}_{i\gamma\sigma} s^{}_{i\gamma\sigma} \bigr)^{-\frac{1}{2}}		
\label{Eq:Hop}
\end{equation}
Here the root operators are identical to one if treated exactly. 
They are introduced to reproduce the non-interacting limit $U = 0$ 
in the mean-field treatment.

To use the slave boson theory, we first diagonalize
the on-site part of $H_{1}$, introducing hole operators 
$h^{\dag}_{\vect{k}\alpha\sigma}$
with $a'_{1g}$ 
and $e'_{g}$ symmetry. In the new basis 
$\big\{ \alpha \big\}$ =  
$\big\{ \ket{a'_{1g}}, \ket{e'_{g1}}, \ket{e'_{g2}} \big\}$,
the Hamiltonian for holes becomes
\begin{align}
	H_{1} = \sum_{\vect{k}} &\sum_{\alpha,\alpha',\sigma} \big( \tilde{\epsilon}_{\alpha\alpha'}(\vect{k}) + \tilde{D}_{\alpha\alpha'} \big) h^{\dag}_{\vect{k}\alpha\sigma} h^{}_{\vect{k}\alpha'\sigma} \notag \\
	\tilde{\epsilon}_{\alpha\alpha'}(\vect{k}) &= P^{-1} \big( \epsilon_{\alpha\alpha'}(\vect{k}) + \epsilon'_{\alpha\alpha'}(\vect{k}) \big) P \notag \\
		&= \begin{pmatrix}
		t_{a}(\vect{k}) & t_{a,1}(\vect{k}) & t_{a,2}(\vect{k}) \\
		t_{a,1}(\vect{k}) & t_{1,1}(\vect{k}) & t_{1,2}(\vect{k}) \\
		t_{a,2}(\vect{k}) & t_{2,1}(\vect{k}) & t_{2,2}(\vect{k})
		\end{pmatrix} \notag \\
	\tilde{D}_{\alpha\alpha'} &= P^{-1} D_{\alpha\alpha'} P
		= D \begin{pmatrix}
		-2 & 0 & 0 \\
		 0 & 1 & 0 \\
		 0 & 0 & 1
		\end{pmatrix} \label{Eq:Hhop2}
\end{align}
with $P$ being the matrix for the change of basis. 
In mean-field the bosonic operators are replaced by numbers, the
values of which are determined self-consistently minimizing the total
energy under the given constraints. 
For a homogeneous paramagnetic phase we have two different 
slave bosons for singly occupied sites, $\moy{s_{ia\sigma}} = s_{a}$ 
for $a'_{1g}$ and $\moy{s_{i\beta\sigma}} = s_{e}$ for $e'_{g\beta}$. 
Eqs.~\eqref{Eq:Cons1} imply
\begin{equation}
	e^{2} + 2s_{a}^{2} + 4s_{e}^{2} = 1; \quad
	\moy{n} = 2s_{a}^{2} + 4s_{e}^{2} = 1-x
\end{equation}
with $\moy{n}$ the total hole occupancy per site. From  Eq.~\eqref{Eq:Hop}, 
the hopping terms become renormalized as $t_{a}(%
\vect{k}) \rightarrow q_{a} t_{a}(\vect{k})$, $t_{e}(\vect{k}) \rightarrow
q_{e} t_{e}(\vect{k})$ and $t_{ae}(\vect{k}) \rightarrow \sqrt{\smash[b]{q_{a}
q_{e}}} t_{ae}(\vect{k})$, where $q_{\alpha }=x/(1-s_{\alpha }^{2})$.

\begin{figure}[htbp]
	\includegraphics[width=.48\linewidth, height=3cm]{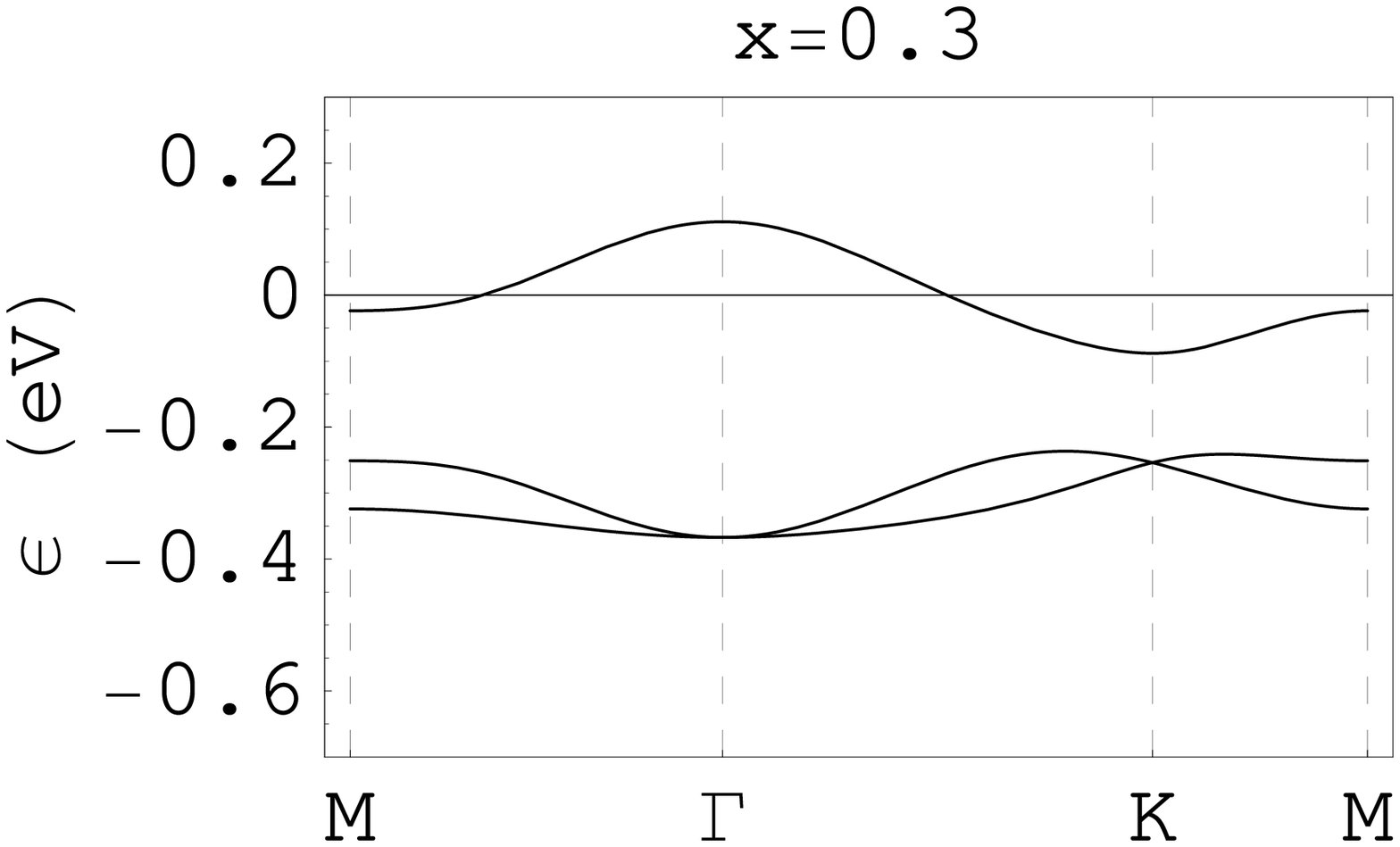} \hfill
	\includegraphics[width=.48\linewidth, height=3cm]{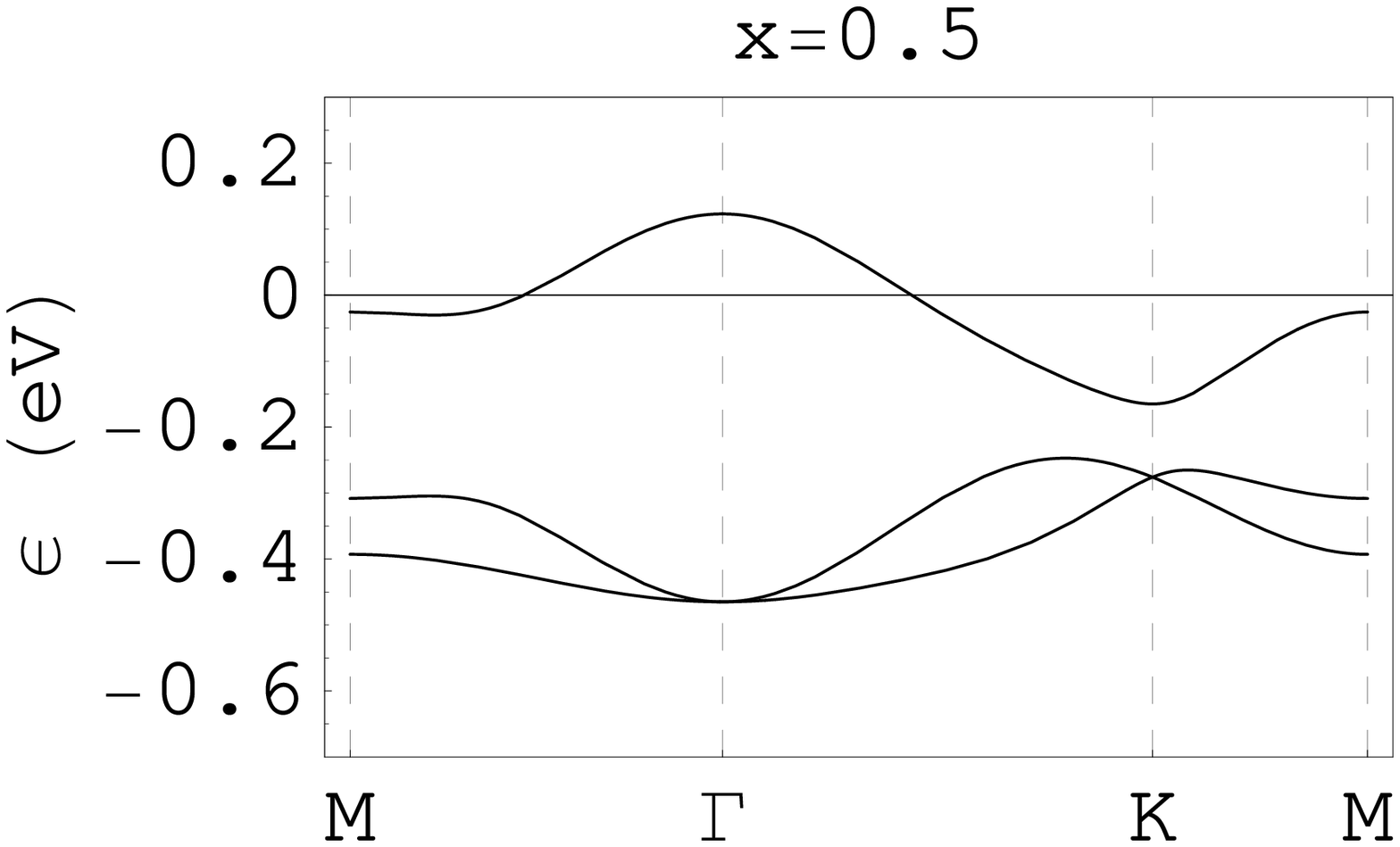}
	\includegraphics[width=.48\linewidth, height=3cm]{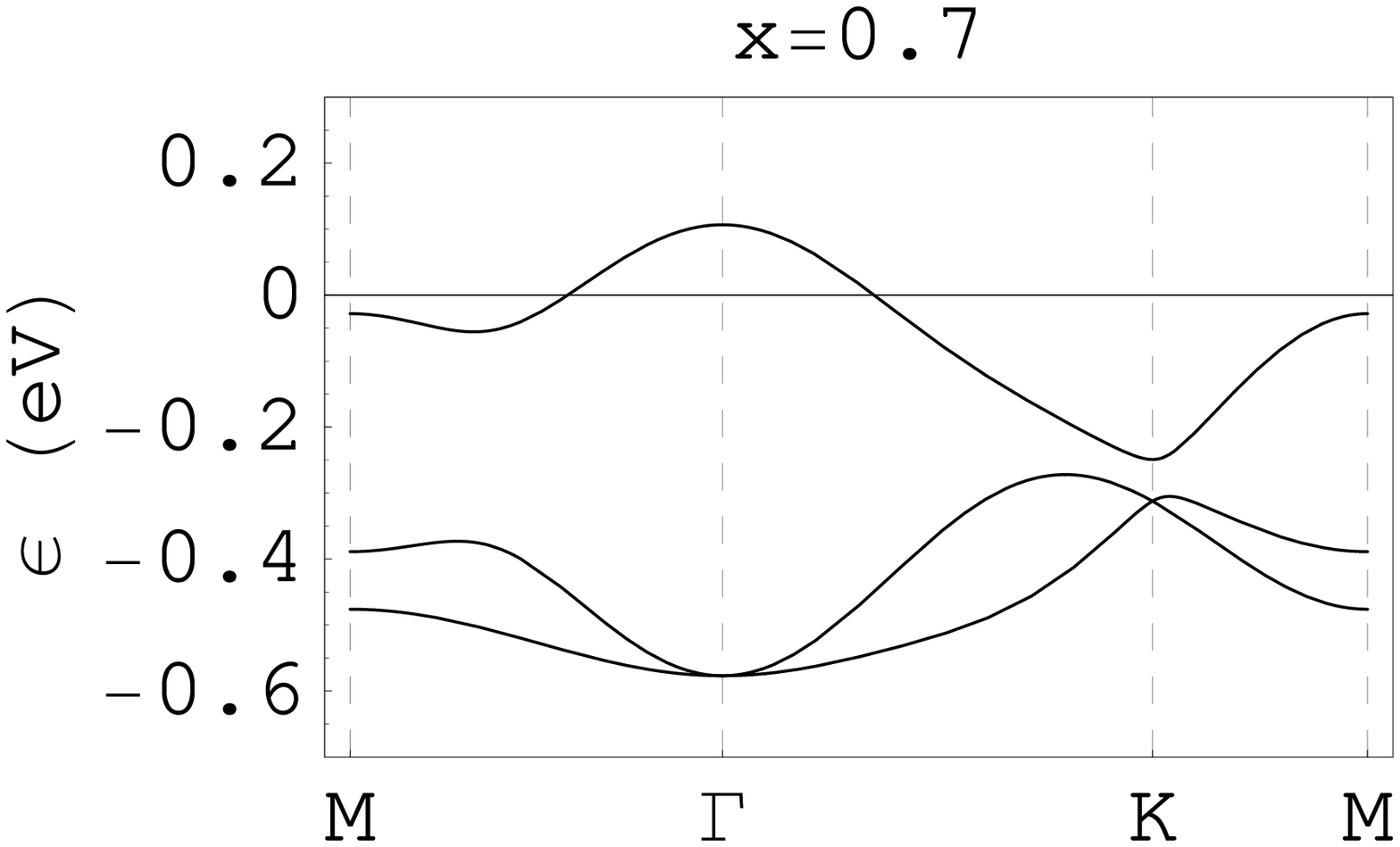} \hfill
	\includegraphics[height=3cm]{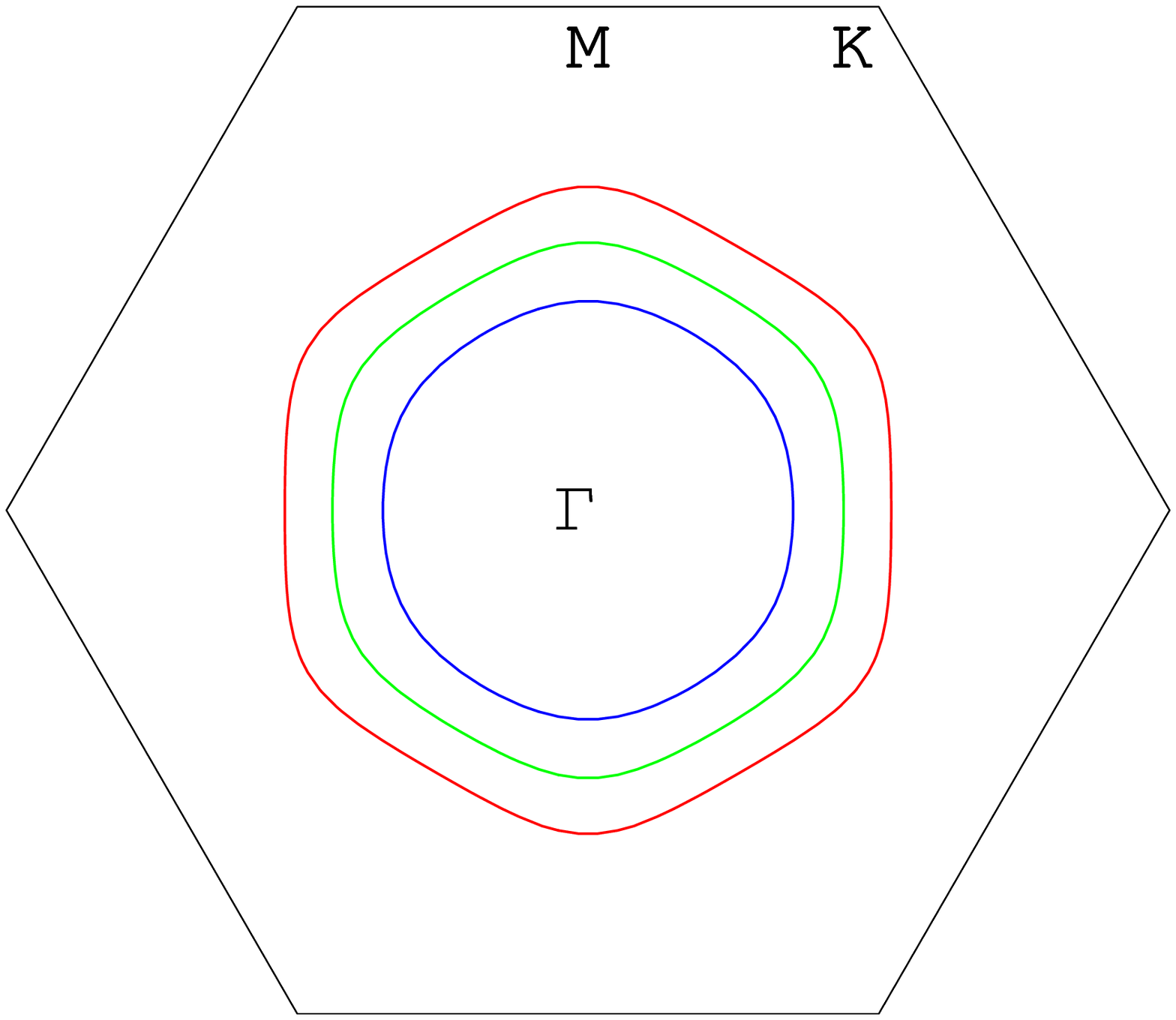}
	\caption{(Color online) Renormalized band-structures for doping $x = 0.3,\ 0.5,\ 0.7$ and corresponding FS in red, green, blue respectively. Larger FS corresponds to lower doping $x$.}
	\label{Fig:Bands}
\end{figure}

In Fig.~\ref{Fig:Bands} we show the resulting dispersion relation for 
electrons (instead of holes to facilitate comparison with previous works) 
for three different values of $x$ near to those measured by ARPES, and the corresponding FS. 
A comparison with the case $U=0$ in the effective model (not shown),
for which the band width is near 0.9 eV, renders it clear 
the effect of the band narrowing as a consequence of the 
remaining correlations in the effective model.
This renormalized bandwidth is stronger as the Mott insulating limit 
$x \rightarrow 0$ is approached. As a consequence of both, 
the relatively strong trigonal splitting due to the large 
Co--O covalency and the band renormalization, 
the hole pockets near the Brillouin zone boundary disappear.
We want to stress that if only one of these effects were present, these pockets would remain. 
In agreement with the most recent ARPES measurements~\cite{PRL95146401}, 
the feature that would give 
rise to hole-pockets near the $K$ points remains under 
$\epsilon_{F}$ with almost no doping dependence 
(while the corresponding band bottom shifts from $\sim -100$~meV between 
$x = 0.3$ and $x = 0.5$). The FS agree with ARPES experiments, showing only a central 
lobe with a hexagonal shape, which is more marked for lower dopings. 
The average amount of $a'_{1g}$ holes on the FS is 38\% weakly 
dependent on doping, the $a'_{1g}$ character is minimum in the 
$\varGamma K$ direction ($\sim 31$\%) and maximum in the $\varGamma M$ 
direction ($\sim 47$\%). This indicates that although only one band 
crosses the $\epsilon_F$, its character is strongly mixed and cannot be 
derived from states of $a'_{1g}$ symmetry only.

In summary, we have derived an effective Hamiltonian $H_{eff}$ to describe 
the low-energy physics of \Na{x}, starting from the ground state of 
CoO$_{6}$ clusters, and calculating the effective hopping between 
different clusters. Previous calculations of the trigonal distortion and
fitting of the XAS experiment using CoO$_{6}$ clusters fixes the 
parameters of $H_{eff}$. This $H_{eff}$ without any new adjustable 
parameter is able to reproduce the main features of the ARPES experiments.

This work was sponsored by PICT 03-12742 of ANPCyT, Argentina, 
ECOS-Sud, France, and DFG KL1842/2, Germany
A.A.A. (T.K.) is  partially supported by CONICET (DAAD).


\begin{thebibliography}{00}

\expandafter\ifx\csname natexlab\endcsname\relax\def\natexlab#1{#1}\fi
\expandafter\ifx\csname bibnamefont\endcsname\relax
  \def\bibnamefont#1{#1}\fi
\expandafter\ifx\csname bibfnamefont\endcsname\relax
  \def\bibfnamefont#1{#1}\fi
\expandafter\ifx\csname citenamefont\endcsname\relax
  \def\citenamefont#1{#1}\fi
\expandafter\ifx\csname url\endcsname\relax
  \def\url#1{\texttt{#1}}\fi
\expandafter\ifx\csname urlprefix\endcsname\relax\def\urlprefix{URL }\fi
\providecommand{\bibinfo}[2]{#2}
\providecommand{\eprint}[2][]{\url{#2}}

\bibitem[{\citenamefont{Terasaki et~al.}(2002)\citenamefont{Terasaki, Tsukada,
  and Iguchi}}]{PRB65195106}
\bibinfo{author}{\bibfnamefont{I.}~\bibnamefont{Terasaki}},
  \bibinfo{author}{\bibfnamefont{I.}~\bibnamefont{Tsukada}}, \bibnamefont{and}
  \bibinfo{author}{\bibfnamefont{Y.}~\bibnamefont{Iguchi}},
  \bibinfo{journal}{Phys. Rev. B} \textbf{\bibinfo{volume}{65}},
  \bibinfo{pages}{195106} (\bibinfo{year}{2002}).

\bibitem[{\citenamefont{Mikami et~al.}(2003)\citenamefont{Mikami, Yoshimura,
  Mori, Sasaki, Funahashi, and Shikano}}]{JJAP427383}
\bibinfo{author}{\bibfnamefont{M.}~\bibnamefont{Mikami}}
  \textit{\bibnamefont{et~al.}},
  \bibinfo{journal}{Jpn. J. Appl. Phys.} \textbf{\bibinfo{volume}{42}},
  \bibinfo{pages}{7383} (\bibinfo{year}{2003}).

\bibitem[{\citenamefont{Takada et~al.}(2003)\citenamefont{Takada, Sakurai,
  Takayama-Muromachi, Izumi, Dilanian, and Sasaki}}]{NAT42253}
\bibinfo{author}{\bibfnamefont{K.}~\bibnamefont{Takada}}
  \textit{\bibnamefont{et~al.}},
  \bibinfo{journal}{Nature} \textbf{\bibinfo{volume}{422}}, \bibinfo{pages}{53}
  (\bibinfo{year}{2003}).

\bibitem[{\citenamefont{Baskaran}(2003)}]{PRL91097003}
\bibinfo{author}{\bibfnamefont{G.}~\bibnamefont{Baskaran}},
  \bibinfo{journal}{Phys. Rev. Lett.} \textbf{\bibinfo{volume}{91}},
  \bibinfo{pages}{097003} (\bibinfo{year}{2003}).

\bibitem[{\citenamefont{Koshibae and Maekawa}(2003)}]{PRL91257003}
\bibinfo{author}{\bibfnamefont{W.}~\bibnamefont{Koshibae}} \bibnamefont{and}
  \bibinfo{author}{\bibfnamefont{S.}~\bibnamefont{Maekawa}},
  \bibinfo{journal}{Phys. Rev. Lett.} \textbf{\bibinfo{volume}{91}},
  \bibinfo{pages}{257003} (\bibinfo{year}{2003}).

\bibitem[{\citenamefont{Indergand et~al.}(2005)\citenamefont{Indergand,
  Yamashita, Kusunose, and Sigrist}}]{PRB71214414}
\bibinfo{author}{\bibfnamefont{M.}~\bibnamefont{Indergand}}
  \textit{\bibnamefont{et~al.}},
  \bibinfo{journal}{Phys. Rev. B} \textbf{\bibinfo{volume}{71}},
  \bibinfo{pages}{214414} (\bibinfo{year}{2005}).

\bibitem[{\citenamefont{Singh}(2000)}]{PRB6113397}
\bibinfo{author}{\bibfnamefont{D.~J.} \bibnamefont{Singh}},
  \bibinfo{journal}{Phys. Rev. B} \textbf{\bibinfo{volume}{61}},
  \bibinfo{pages}{13397} (\bibinfo{year}{2000}).

\bibitem[{\citenamefont{Lee et~al.}(2004)\citenamefont{Lee, Kunes, and
  Pickett}}]{PRB70045104}
\bibinfo{author}{\bibfnamefont{K.~W.} \bibnamefont{Lee}},
  \bibinfo{author}{\bibfnamefont{J.}~\bibnamefont{Kunes}}, \bibnamefont{and}
  \bibinfo{author}{\bibfnamefont{W.~E.} \bibnamefont{Pickett}},
  \bibinfo{journal}{Phys. Rev. B} \textbf{\bibinfo{volume}{70}},
  \bibinfo{pages}{045104} (\bibinfo{year}{2004}).

\bibitem[{\citenamefont{Hasan et~al.}(2004)\citenamefont{Hasan, Chuang, Qian,
  Li, Kong, Kuprin, Fedorov, Kimmerling, Rotenberg, Rossnagel
  et~al.}}]{PRL92246402}
\bibinfo{author}{\bibfnamefont{M.~Z.} \bibnamefont{Hasan}}
  \textit{\bibnamefont{et~al.}},
  \bibinfo{journal}{Phys. Rev. Lett.}
  \textbf{\bibinfo{volume}{92}}, \bibinfo{pages}{246402}
  (\bibinfo{year}{2004}).

\bibitem[{\citenamefont{Yang et~al.}(2004)\citenamefont{Yang, Wang, Sekharan,
  Matsui, Souma, Sato, Takahashi, Takeuchi, Campuzano, Jin
  et~al.}}]{PRL92246403}
\bibinfo{author}{\bibfnamefont{H.~B.} \bibnamefont{Yang}}
  \textit{\bibnamefont{et~al.}},
  \bibinfo{journal}{Phys. Rev. Lett.} \textbf{\bibinfo{volume}{92}},
  \bibinfo{pages}{246403} (\bibinfo{year}{2004}).

\bibitem[{\citenamefont{Yang et~al.}(2005)\citenamefont{Yang, Pan, Sekharan,
  Sato, Souma, Takahashi, Jin, Sales, Mandrus, Fedorov et~al.}}]{PRL95146401}
\bibinfo{author}{\bibfnamefont{H.~B.} \bibnamefont{Yang}}
  \textit{\bibnamefont{et~al.}},
  \bibinfo{journal}{Phys. Rev. Lett.}
  \textbf{\bibinfo{volume}{95}}, \bibinfo{pages}{146401}
  (\bibinfo{year}{2005}).

\bibitem[{\citenamefont{Wu et~al.}(2005)\citenamefont{Wu, Huang, Okamoto,
  Tanaka, Lin, Chou, Fujimori, and Chen}}]{PRL94146402}
\bibinfo{author}{\bibfnamefont{W.~B.} \bibnamefont{Wu}}
  \textit{\bibnamefont{et~al.}},
  \bibinfo{journal}{Phys. Rev. Lett.} \textbf{\bibinfo{volume}{94}},
  \bibinfo{pages}{146402} (\bibinfo{year}{2005}).

\bibitem[{\citenamefont{Kroll et~al.}({\natexlab{a}})\citenamefont{Kroll,
  Knupfer, Geck, Hess, Schwieger, Krabbes, Sekar, Batchelor, Berger, and
  B\"uchner}}]{PRB74115123}
\bibinfo{author}{\bibfnamefont{T.}~\bibnamefont{Kroll}}
  \textit{\bibnamefont{et~al.}},
  \bibinfo{journal}{Phys. Rev. B} \textbf{\bibinfo{volume}{74}},
  \bibinfo{pages}{115123} (\bibinfo{year}{2006}).

\bibitem[{\citenamefont{Kroll et~al.}({\natexlab{b}})\citenamefont{Kroll,
  Aligia, and Sawatzky}}]{PRB74115124}
\bibinfo{author}{\bibfnamefont{T.}~\bibnamefont{Kroll}},
  \bibinfo{author}{\bibfnamefont{A.~A.} \bibnamefont{Aligia}},
  \bibnamefont{and} \bibinfo{author}{\bibfnamefont{G.}~\bibnamefont{Sawatzky}},
  \bibinfo{journal}{Phys. Rev. B} \textbf{\bibinfo{volume}{74}},
  \bibinfo{pages}{115124} (\bibinfo{year}{2006}).

\bibitem[{\citenamefont{Landron and Lepetit}(2006)}]{CM0605454}
\bibinfo{author}{\bibfnamefont{S.}~\bibnamefont{Landron}} \bibnamefont{and}
  \bibinfo{author}{\bibfnamefont{M.~B.} \bibnamefont{Lepetit}}, 
  \bibinfo{note}{cond-mat/0605454}.

\bibitem[{\citenamefont{Zhou et~al.}(2005)\citenamefont{Zhou, Gao, Ding, Lee,
  and Wang}}]{PRL94206401}
\bibinfo{author}{\bibfnamefont{S.}~\bibnamefont{Zhou}}
  \textit{\bibnamefont{et~al.}},
  \bibinfo{journal}{Phys. Rev. Lett.} \textbf{\bibinfo{volume}{94}},
  \bibinfo{pages}{206401} (\bibinfo{year}{2005}).

\bibitem[{\citenamefont{Zhang et~al.}(2004)\citenamefont{Zhang, Luo, Cohen, and
  Louie}}]{PRL93236402}
\bibinfo{author}{\bibfnamefont{P.}~\bibnamefont{Zhang}}
  \textit{\bibnamefont{et~al.}},
  \bibinfo{journal}{Phys. Rev. Lett.} \textbf{\bibinfo{volume}{93}},
  \bibinfo{pages}{236402} (\bibinfo{year}{2004}).

\bibitem[{\citenamefont{Ishida et~al.}(2005)\citenamefont{Ishida, Johannes, and
  Liebsch}}]{PRL94196401}
\bibinfo{author}{\bibfnamefont{H.}~\bibnamefont{Ishida}},
  \bibinfo{author}{\bibfnamefont{M.~D.} \bibnamefont{Johannes}},
  \bibnamefont{and} \bibinfo{author}{\bibfnamefont{A.}~\bibnamefont{Liebsch}},
  \bibinfo{journal}{Phys. Rev. Lett.} \textbf{\bibinfo{volume}{94}},
  \bibinfo{pages}{196401} (\bibinfo{year}{2005}).

\bibitem[{\citenamefont{Singh and Kasinathan}(2006)}]{CM0604002}
\bibinfo{author}{\bibfnamefont{D.~J.} \bibnamefont{Singh}} \bibnamefont{and}
  \bibinfo{author}{\bibfnamefont{D.}~\bibnamefont{Kasinathan}}, 
  \bibinfo{note}{cond-mat/0604002}.

\bibitem[{\citenamefont{Zhang}(1989)}]{PRB397375}
\bibinfo{author}{\bibfnamefont{F.~C.} \bibnamefont{Zhang}},
  \bibinfo{journal}{Phys. Rev. B} \textbf{\bibinfo{volume}{39}},
  \bibinfo{pages}{7375} (\bibinfo{year}{1989}).

\bibitem[{\citenamefont{Aligia et~al.}(1994)\citenamefont{Aligia, Simon, and
  Batista}}]{PRB4913061}
\bibinfo{author}{\bibfnamefont{A.~A.} \bibnamefont{Aligia}},
  \bibinfo{author}{\bibfnamefont{M.~E.} \bibnamefont{Simon}}, \bibnamefont{and}
  \bibinfo{author}{\bibfnamefont{C.~D.} \bibnamefont{Batista}},
  \bibinfo{journal}{Phys. Rev. B} \textbf{\bibinfo{volume}{49}},
  \bibinfo{pages}{13061} (\bibinfo{year}{1994}).

\bibitem[{\citenamefont{Garc\'ia et~al.}(1995)\citenamefont{Garc\'ia, Castell,
  Caballol, and Malrieu}}]{CPL238222}
\bibinfo{author}{\bibfnamefont{V.~M.} \bibnamefont{Garc\'ia}}
  \textit{\bibnamefont{et~al.}},
  \bibinfo{journal}{Chem. Phys. Lett.} \textbf{\bibinfo{volume}{238}},
  \bibinfo{pages}{222} (\bibinfo{year}{1995}).

\bibitem[{\citenamefont{Kotliar and Ruckenstein}(1986)}]{PRL571362}
\bibinfo{author}{\bibfnamefont{G.}~\bibnamefont{Kotliar}} \bibnamefont{and}
  \bibinfo{author}{\bibfnamefont{A.~E.} \bibnamefont{Ruckenstein}},
  \bibinfo{journal}{Phys. Rev. Lett.} \textbf{\bibinfo{volume}{57}},
  \bibinfo{pages}{1362} (\bibinfo{year}{1986}).

\bibitem[{\citenamefont{Fr\'esard and Kotliar}(1997)}]{PRB5612909JPSJ661391PRB475095}
\bibinfo{author}{\bibfnamefont{R.}~\bibnamefont{Fr\'esard}} \bibnamefont{and}
  \bibinfo{author}{\bibfnamefont{G.}~\bibnamefont{Kotliar}},
  \bibinfo{journal}{Phys. Rev. B} \textbf{\bibinfo{volume}{56}},
  \bibinfo{pages}{12909} (\bibinfo{year}{1997}); 
\bibinfo{author}{\bibfnamefont{H.}~\bibnamefont{Hasegawa}},
  \bibinfo{journal}{J. Phys. Soc. Jpn.} \textbf{\bibinfo{volume}{66}},
  \bibinfo{pages}{1391} (\bibinfo{year}{1997}); 
\bibinfo{author}{\bibfnamefont{V.}~\bibnamefont{Dorin}} \bibnamefont{and}
  \bibinfo{author}{\bibfnamefont{P.}~\bibnamefont{Schlottmann}},
  \bibinfo{journal}{Phys. Rev. B} \textbf{\bibinfo{volume}{47}},
  \bibinfo{pages}{5095} (\bibinfo{year}{1993}).

\bibitem[{\citenamefont{Feiner et~al.}(1996)\citenamefont{Feiner, Jefferson,
  and Raimondi}}]{PRB538751}
\bibinfo{author}{\bibfnamefont{L.~F.} \bibnamefont{Feiner}},
  \bibinfo{author}{\bibfnamefont{J.~H.} \bibnamefont{Jefferson}},
  \bibnamefont{and} \bibinfo{author}{\bibfnamefont{R.}~\bibnamefont{Raimondi}},
  \bibinfo{journal}{Phys. Rev. B} \textbf{\bibinfo{volume}{53}},
  \bibinfo{pages}{8751} (\bibinfo{year}{1996}), \bibinfo{note}{references
  therein}.

\bibitem[{\citenamefont{Simon et~al.}(1997)\citenamefont{Simon, Aligia, and
  Gagliano}}]{PRB565637}
\bibinfo{author}{\bibfnamefont{M.~E.} \bibnamefont{Simon}},
  \bibinfo{author}{\bibfnamefont{A.~A.} \bibnamefont{Aligia}},
  \bibnamefont{and} \bibinfo{author}{\bibfnamefont{E.~R.}
  \bibnamefont{Gagliano}}, \bibinfo{journal}{Phys. Rev. B}
  \textbf{\bibinfo{volume}{56}}, \bibinfo{pages}{5637} (\bibinfo{year}{1997}).

\bibitem[{\citenamefont{Petrone and Aligia}(2002)}]{PRB66104418}
\bibinfo{author}{\bibfnamefont{P.}~\bibnamefont{Petrone}} \bibnamefont{and}
  \bibinfo{author}{\bibfnamefont{A.~A.} \bibnamefont{Aligia}},
  \bibinfo{journal}{Phys. Rev. B} \textbf{\bibinfo{volume}{66}},
  \bibinfo{pages}{104418} (\bibinfo{year}{2002}).

\end{thebibliography}
\end{document}